\documentclass[12pt,preprint]{aastex}

\begin{document}

\title{The Prompt Inventory from Very Massive Stars and Elemental
Abundances in Ly$\alpha$ Systems}
\author{Y.-Z. Qian\altaffilmark{1}, W. L. W. Sargent\altaffilmark{2},
and G. J. Wasserburg\altaffilmark{3}}
\altaffiltext{1}{School of
Physics and Astronomy, University of Minnesota, Minneapolis, MN
55455; qian@physics.umn.edu.}
\altaffiltext{2}{Department of Astronomy,
California Institute of Technology, Pasadena, CA 91125}
\altaffiltext{3}{The Lunatic Asylum, Division of
Geological and Planetary Sciences, California Institute of
Technology, Pasadena, CA 91125.}

\begin{abstract}
It has been proposed that very massive stars (VMSs) dominated
heavy element production until a ``metallicity'' threshold
corresponding to [Fe/H]~$\approx -3$ was reached. This results 
in a prompt ($P$) inventory of elements, the abundances of which
were determined from observations of Galactic halo stars with
[Fe/H]~$\approx -3$. We calculate $\Omega_P({\rm E})$ from
the $P$ inventory for
a large number of elements in the IGM.
Using the available data on $\Omega({\rm E}_{\rm ion})$ for
\ion{C}{4}, \ion{O}{6}, and \ion{Si}{4} in Ly$\alpha$ systems, 
we find that the ionization 
fractions calculated from 
$\Omega({\rm E}_{\rm ion})/\Omega_P({\rm E})$ are, within 
reasonable uncertainties, compatible 
with the values estimated from ionization models
for Ly$\alpha$ systems. This agreement appears to hold from
$z\sim 0.09$ to $\sim 4.6$, indicating that the bulk of
the baryonic matter remains dispersed with a fixed chemical
composition. We conclude 
that the $P$ inventory was established in the epoch prior to 
$z\sim 4.6$. The dispersal of processed baryonic matter to the 
general IGM is considered to be the
result of energetic VMS explosions that disrupted baryonic
aggregates until the ``metallicity'' threshold was reached to
permit normal astration. The formation of most
galaxies is considered to have occurred subsequent to the
achievement of this metallicity threshold in the IGM.
\end{abstract}

\section{Introduction}
We present predictions for the abundances of a large number of
elements in the intergalactic medium (IGM), especially the
Ly$\alpha$ systems, based on considerations of a prompt ($P$)
inventory that is dominated by the contributions from very massive
($\gtrsim 100\,M_{\odot}$) stars (VMSs). 
A comparison will be made between the predictions and the
available observational data. The model of the $P$ inventory
(Wasserburg \& Qian 2000a; Qian \& Wasserburg 2001a; Qian \&
Wasserburg 2002) claims that, prior to the achievement of the
condition corresponding to a metallicity of [Fe/H]~$\approx -3$ in
the IGM, the predominant mechanism of astration is by formation of
VMSs. This
condition is identified based on the sharp rise in the abundances
of the heavy $r$-process elements such as Ba and Eu in Galactic
halo stars with [Fe/H]~$\approx -3$ (e.g., McWilliam et al. 1995;
Burris et al. 2000). This sharp rise is considered to represent
the occurrence of Type II supernovae (SNe II)
that result from evolution of
normal stars with masses of $\sim (10$--$60)\,M_{\odot}$ in the
absence of VMSs. Further support for a threshold at [Fe/H]~$\approx
-3$ for normal astration comes from the observation that all
damped Ly$\alpha$ systems with redshift $z\approx 1.5$--4.5 have a
minimum [Fe/H] of $\approx -2.7$ (e.g., Prochaska \& Wolfe 2000;
Prochaska, Gawiser, \& Wolfe 2001). The metallicity of
[Fe/H]~$\approx -3$ is considered to represent the critical
condition for sufficient cooling and fragmentation of gas clouds
to permit major formation of normal stars with masses of $\sim
(1$--$60)\,M_{\odot}$ (Wasserburg \& Qian 2000a). This is supported
by ab initio calculations of Bromm et al. (2001), which showed
that a critical metallicity of $\sim 5\times 10^{-4}$ the solar 
value (corresponding to [Fe/H]~$\sim -3$) is
required for significant formation of low-mass protostellar
aggregates.

It was found that the observed elemental abundances in stars with
$-3\lesssim{\rm [Fe/H]}< -1$ could be quantitatively explained by
the contributions from: (1) the $P$ inventory at [Fe/H]~$\approx
-3$ for all the elements up to Zr; (2) a high-frequency type of
SNe II [SNe II($H$)] that are responsible for the heavy
$r$-process elements and some light $r$-process elements below Ba
but produce very little of the elements from N to Zn including Fe;
and (3) a low-frequency type of SNe II [SNe II($L$)] that are
responsible for the bulk of the light $r$-process elements and the
part of the elements from N to Zn attributable to SNe II (Qian \&
Wasserburg 2001b, 2002). It was further found by reducing the
observational data on metal-poor Galactic halo stars that the yield
patterns of SNe II($L$) and VMS are almost the same for all the
elements below Sr except for some small but significant shifts at
Na, Al, V, Cr, Mn, and Co. Thus, the abundance ratios of the
so-called $\alpha$ elements relative to Fe such as Mg/Fe, Si/Fe,
Ca/Fe, and Ti/Fe remain essentially constant at $-4\lesssim{\rm
[Fe/H]}< -1$ although the dominant source of the $\alpha$ elements
and Fe changes from VMSs in the regime of [Fe/H]~$<-3$ to SNe II($L$)
in the regime of $-3<{\rm [Fe/H]}<-1$. These ratios are all greater
than the corresponding solar values by a factor of $\sim 3$
because $\sim 2/3$ of the solar Fe inventory is from Type Ia
Supernovae, which later added the Fe group elements but little of
the $\alpha$ elements at [Fe/H]~$\gtrsim -1$. This simple
explanation should serve to clarify the so-called
``overproduction'' of what are referred to as the $\alpha$
elements at low metallicities.

It was considered that the $P$ inventory at [Fe/H]~$\approx -3$ is
dominated by the contributions from VMSs with small contributions
from SNe II. The rare occurrence of Galactic
halo stars with $-4\lesssim {\rm [Fe/H]}<-3$, 
which must have masses of $\sim 1\,M_\odot$ in order to
survive to the present epoch,
is considered to represent concomitant formation of some 
``normal'' stars in
the regime where VMSs dominated astration (Qian \& Wasserburg
2002). It was also suggested that VMSs were the source for
reionization of the IGM (e.g., Qian \& Wasserburg 2001b). A recent
study by Oh et al. (2001) showed that in producing a ``metal''
inventory corresponding to [Fe/H]~$\sim -3$ (or [Si/H]~$\sim
-2.5$), VMSs also provide $\sim 1$--10 ionizing photons for H and
He. This was estimated from the VMS model yields calculated by
Heger \& Woosley (2002) and the VMS emission spectra calculated by
Bromm, Kudritzki, \& Loeb (2001). It appears that by processing
$\sim 10^{-5}$--$10^{-4}$ of the baryonic matter, VMSs are capable
of providing the $P$ inventory and also of reionizing the IGM
prior to the formation of most galaxies.

Further justification for the role of VMSs is found from the very
low Ba abundances at [Fe/H]~$< -3$. This requires that any Ba
produced by SNe II at [Fe/H]~$<-3$ had to be mixed in a dilution
mass of $\sim (10^6$--$10^7)\,M_{\odot}$, to be compared with the
dilution mass of $\sim 3\times 10^4\,M_{\odot}$ for SNe II at
[Fe/H]~$> -3$. It is considered that the severely disruptive VMS
activities at [Fe/H]~$<-3$ would not often permit the preservation
of baryonic aggregates with masses of $\lesssim
(10^6$--$10^7)\,M_{\odot}$ and that the $P$ inventory at
[Fe/H]~$\approx -3$ should represent the chemical composition of
the general IGM (Qian \& Wasserburg 2002). Elemental abundances in
the IGM are best measured from absorption lines produced by regions
of enhanced density such as Ly$\alpha$ systems that are
illuminated by Quasi-Stellar Objects (QSOs). For a given
QSO, absorption lines occur for narrow
ranges in redshift (below $z$ of the QSO) and the
absorbing systems represent concentrations of dispersed baryonic
matter between the QSO and the observer. 
Extensive studies have determined the
abundances of \ion{C}{4}, \ion{O}{6}, and
\ion{Si}{4} in Ly$\alpha$ systems with a wide
range in redshift from $z\sim 0.09$ up to $z\sim 5.3$. In the following,
we will present the abundances of a large number of elements in
the IGM based on the model of the $P$ inventory (\S2) and compare
these results with the available data on Ly$\alpha$ systems (\S3).
The cosmological implications of our model will be discussed (\S4).

\section{Elemental Abundances in Ly$\alpha$ Systems as Determined
from the $P$ Inventory}
The abundances for the $P$ inventory are given in Table 1
in the standard spectroscopic notation as
\begin{equation}
\log\epsilon_P({\rm E})=\log\left({{\rm E}\over {\rm H}}\right)_P+12,
\label{pinv}
\end{equation}
where (E/H)$_P$ is the number ratio of an element E relative to
hydrogen for the $P$ inventory. The $\log\epsilon_P({\rm E})$
values in Table 1 are directly obtained from observational data on
Galactic halo stars with [Fe/H]~$\approx -3$ and correspond to
those given in Qian \& Wasserburg (2002). The only element for
which there may be a serious uncertainty is O. The 
$\log\epsilon_P({\rm O})$ value
assumed here is based on observations of HD 115444 and
HD 122563 (Westin et al. 2000). Other extensive
studies (e.g., Israelian, Garc\'ia L\'opez, \& Rebolo 1998;
Boesgaard et al. 1999) indicate that
$\log\epsilon_P({\rm O})$ may be up to $\sim 0.3$ dex higher than
the value given in Table 1. The precise value of
$\log\epsilon_P({\rm O})$ is an important issue that remains to be
resolved.

We consider that essentially all of the baryonic matter was
ionized and dispersed in the IGM when VMS activities ceased and
that the chemical composition of this baryonic matter was
represented by the $P$ inventory. The mechanism for maintaining the
ionized state of the IGM will not be discussed here. We
assume that subsequent to the acquisition of the $P$ inventory,
the bulk of the baryonic matter remains dispersed in the form of
Ly$\alpha$ systems. We further assume that these systems have
experienced no further evolution in elemental abundances and that
they represent regions of enhanced density comoving with the
expansion of the universe. Under these assumptions, the number
ratio of an element E relative to H in the Ly$\alpha$ systems is
the same as that for the $P$ inventory and can
be written as
\begin{equation}
\left({{\rm E}\over{\rm H}}\right)_P={\Omega_P({\rm
E})/m_{\rm E}\over\Omega({\rm H})/m_{\rm H}}={\Omega_P({\rm
E})/A_{\rm E}\over\Omega({\rm H})}, \label{omega}
\end{equation}
where $\Omega_P({\rm E})$ and $\Omega({\rm H})$ are the fractions of
the critical mass density $\rho_{\rm cri}=3H_0^2/(8\pi G)$
contributed by E and H, respectively, at the present epoch,
$m_{\rm E}$ and $m_{\rm H}$ are the atomic masses of E and H,
respectively, and $A_{\rm E}=m_{\rm E}/m_{\rm H}$ is the
atomic mass number of E relative to hydrogen. The symbols $H_0$
and $G$ in the expression of $\rho_{\rm cri}$ are the Hubble and
gravitational constants, respectively. The baryonic contribution
to $\rho_{\rm cri}$ is found to be $\Omega_b\approx 0.02h^{-2}$
based on measurements of deuterium abundances in QSO absorption
systems (e.g., Burles, Nollett, \& Turner 2001) and 
the power spectrum of the
cosmic microwave background (e.g., de Bernardis et al. 2002),
where $h$ is $H_0$ in units of 100 km s$^{-1}$ Mpc$^{-1}$. This
gives $\Omega({\rm H})\approx X_{\rm H}\Omega_b\approx
0.0152h^{-2}$, where $X_{\rm H}\approx 0.76$ is the mass fraction
of H. Combining equations (\ref{pinv}) and (\ref{omega}), we
obtain
\begin{equation}
\Omega_P({\rm E})=\Omega({\rm H})A_{\rm E}\times
10^{\log\epsilon_P({\rm E})-12} \approx 1.52h^{-2}A_{\rm E}\times
10^{\log\epsilon_P({\rm E})-14}. \label{omege}
\end{equation}
Note that $\Omega_P({\rm E})$ does not explicitly depend on 
cosmological models except for the numerical values of $H_0$ and 
$\Omega_b$. The values of $\Omega_P({\rm E})$ calculated from the $P$
inventory are given in Table 1 for $h=0.65$.  At
present, data are available only for several elements such as C,
O, and Si in Ly$\alpha$ systems.

\section{Comparison with Observational Data}
Absorption lines produced by ions such as \ion{C}{4}, \ion{O}{6}, and
\ion{Si}{4} in regions of enhanced density such as the Ly$\alpha$ 
systems have been measured in the
spectra of QSOs that illuminate such regions.
The column density of the ions of an element E, E$_{\rm ion}$, in
a redshift interval $z_-<z<z_+$ is
\begin{equation}
N({\rm E}_{\rm ion})=c\int_{z_-}^{z_+}{n({\rm E}_{\rm ion},z)\over
(1+z)H(z)}dz,
\end{equation}
where $c$ is the speed of light, $n({\rm E}_{\rm ion},z)$ is the
number density of E$_{\rm ion}$ at redshift $z$, and $H(z)$ is the
Hubble parameter at $z$. Under the assumptions
that the bulk of the baryonic matter resides in Ly$\alpha$ systems
and that these systems experience no evolution other than comoving
with the expansion of the universe, we obtain
\begin{equation}
N({\rm E}_{\rm ion})={cn_0({\rm E}_{\rm ion})\over H_0}
\int_{z_-}^{z_+}{(1+z)^2\over H(z)/H_0}dz,
\label{nion}
\end{equation}
where $n_0({\rm E}_{\rm ion})=n({\rm E}_{\rm ion},z)/(1+z)^3$ is
the number density of E$_{\rm ion}$ at the present epoch.
By defining (Bahcall \& Peebles 1969)
\begin{equation}
X(z)\equiv\int_0^z{(1+z')^2\over H(z')/H_0}dz'
\label{xz}
\end{equation}
and $\Delta X\equiv X(z_+)-X(z_-)$, equation (\ref{nion}) can be
rewritten as
\begin{equation}
N({\rm E}_{\rm ion})={cn_0({\rm E}_{\rm ion})\over H_0}\Delta X.
\label{ndx}
\end{equation}
If absorption lines due to E$_{\rm ion}$ have been
measured for a statistical sample of Ly$\alpha$ systems at different 
redshift along the lines of sight for a number of QSOs,
the terms
$N({\rm E}_{\rm ion})$ and $\Delta X$ in equation (\ref{ndx}) are
replaced by the sums $\sum N({\rm E}_{\rm ion})$ and $\sum \Delta
X$ over the sampled region. It follows that the fraction of
$\rho_{\rm cri}$ contributed by E$_{\rm ion}$ at the present epoch,
$\Omega({\rm E}_{\rm ion})=n_0({\rm E}_{\rm ion})m_{\rm E}/
\rho_{\rm cri}$, is
\begin{equation}
\Omega({\rm E}_{\rm ion})=
{H_0\over c\rho_{\rm cri}}{\sum N({\rm E}_{\rm ion})
\over\sum \Delta X}m_{\rm E}=9.55\times 10^{-24}h^{-1}A_{\rm E}
{\sum N({\rm E}_{\rm ion})\over\sum \Delta X},
\label{omegi}
\end{equation}
where the last equality is obtained for $\sum N({\rm E}_{\rm ion})$
in units of cm$^{-2}$. If the
model of the $P$ inventory is correct, the value of $\Omega({\rm
E}_{\rm ion})$ obtained from equation (\ref{omegi}) is related to
that of $\Omega_P({\rm E})$ from equation (\ref{omege}) through the
ionization fraction $f({\rm E}_{\rm ion})=\Omega({\rm E}_{\rm
ion})/\Omega_P({\rm E})$.

The calculation of $\Omega({\rm E}_{\rm ion})$ through 
$\Delta X$ requires a specific cosmological
model. This is distinct from the more direct calculation of
$\Omega_P({\rm E})$. The choice of a particular model makes little
difference for $z\ll 1$, as all models give $\Delta X\approx
z_+-z_-$ in this limit. For simplicity in considering the data
on Ly$\alpha$ systems with a wide range in $z$, we adopt a
flat cosmological model with $\Omega_m+\Omega_\Lambda=1$, where
$\Omega_m$ and $\Omega_\Lambda$ are the fractions of $\rho_{\rm
cri}$ contributed by matter (baryonic and nonbaryonic) and the
cosmological constant, respectively, at the present epoch. With
this model, $[H(z)/H_0]^2=\Omega_m(1+z)^3+\Omega_\Lambda$ and
the expression of $X(z)$ from equation (\ref{xz}) is
\begin{equation}
X(z)={2\over
3\Omega_m}\left[\sqrt{\Omega_m(1+z)^3+1-\Omega_m}-1\right],
\end{equation}
which gives $\Delta X \approx z_+ - z_-$ for $z \ll 1$ and $\Delta
X \approx {2\over 3}\Omega^{-1/2}_m\left[(1 + z_+)^{3/2} - (1 +
z_-)^{3/2}\right]$ for $1 + z\gg [(1-\Omega_m)/\Omega_m]^{1/3}$.

A comparison of the observed values of $\Omega({\rm E}_{\rm ion})$
with the values corresponding to the $P$ inventory requires that 
the same cosmological model and parameters be used. The data on
Ly$\alpha$ systems with a wide range in $z$ as reported
in the literature were based on
somewhat different cosmological models and parameters.
We have recalculated $\Omega({\rm E}_{\rm ion})$ from the published
data so that the recalculated results correspond to the flat
cosmological model adopted here with $\Omega_m = 0.3$,
$\Omega_{\Lambda} = 0.7$, and $h = 0.65$.
The data on \ion{C}{4} and \ion{Si}{4} are taken from  
Songaila (2001) and those on \ion{O}{6} from Burles \& Tytler (1996), 
Tripp, Savage, \& Jenkins (2000), Savage et al. (2002), and Simcoe, 
Sargent, \& Rauch (2002; no recalculation necessary).
The recalculated $\Omega({\rm E}_{\rm ion})$ values are given in
Table 2. Ideally, the fraction $f({\rm E}_{\rm ion})$ of 
the element E that is ionized to the particular state observed can
be calculated for the Ly$\alpha$ systems from which
$\Omega({\rm E}_{\rm ion})$ is obtained. A comparison can then be
made of $\Omega({\rm E})=\Omega({\rm E}_{\rm ion})/
f({\rm E}_{\rm ion})$ with $\Omega_P({\rm E})$ calculated from
the $P$ inventory. As values of $f({\rm E}_{\rm ion})$ are not always 
well known, we have tabulated the ratio $\Omega({\rm E}_{\rm
ion})/\Omega_P({\rm E})$ for \ion{C}{4}, \ion{O}{6}, and
\ion{Si}{4} in Table 2. This ratio should correspond to 
$f({\rm E}_{\rm ion})$ if the model of the $P$ inventory is valid. The
values of $\Omega({\rm E}_{\rm ion})/\Omega_P({\rm E})$ in Table 2
are in reasonable agreement with $f(\mbox{\ion{C}{4}})\sim 0.5$,
$f(\mbox{\ion{O}{6}})\sim 0.2$, and $f(\mbox{\ion{Si}{4}})~\sim 0.1$
as estimated from various ionization models for Ly$\alpha$ systems. 

For all the data sets used in the above comparison, there may be 
significant uncertainties in
estimates of both $\sum N({\rm E}_{\rm ion})$ and $\sum\Delta X$ due
to sampling problems. The mechanisms of ionization are also not
well understood. Considering the very different and complex data
sets used and the extreme simplicity of our model for the $P$ 
inventory, we conclude that there is remarkable agreement between
the observations and the predictions for the abundances in the IGM.
We note that the apparently high values of (O/H)/(O/H)$_\odot$
and (C/H)/(C/H)$_\odot$ reported by Burles \& Tytler (1996)
are off due to the omission of a factor
$1/A_{\rm E}$ in their equation for $\zeta(z)$ (Tytler 2002, personal
communication). We also note that the values of [O/H] used by
Tripp et al. (2000) and Savage et al. (2002) in
estimating the lower bounds to $\Omega_b$ 
may not be appropriate in view of the results presented here.
We consider that a more
detailed comparison of the model for the $P$ inventory and the
observations must await the availability of more complete data
sets with appropriate estimates of errors including those in the
term $\sum \Delta X$. It would be most useful to theoretical studies 
of Ly$\alpha$ systems if observational reports included explicit 
statements on the
value of $\sum N({\rm E}_{\rm ion})$ and on the detailed redshift
intervals for calculation of $\sum\Delta X$ with corrections for
the obscured part of the redshift path. We note that compared with
the flat cosmological model with $\Omega_m = 0.3$ and 
$\Omega_{\Lambda}= 0.7$ as adopted here, a flat model with
$\Omega_m = 1$ (i.e., with the deceleration parameter $q_0 = 0.5$) 
gives smaller $\Delta X$ values, and hence larger 
$\Omega({\rm E}_{\rm ion})$ values, by a factor of 1.83 at high $z$.
It is possible that the model for the $P$ inventory
may be of use in evaluating various cosmological models if
the ionization mechanism for Ly$\alpha$ systems is better understood.

\section{Discussion}
The results in Table 2 strongly support our assumptions that
subsequent to the acquisition of the $P$ inventory, the bulk of
baryonic matter remains dispersed in the form of Ly$\alpha$
systems and that these systems have experienced little or no
chemical evolution but are simply comoving with the expansion of
the universe. It is striking to note that these assumptions hold
from $z\sim 2.5$ down to $\sim 0.09$ based on the \ion{O}{6} data
and from $z\sim 4.6$ down to $\sim 2$ based on the \ion{C}{4} and
\ion{Si}{4} data. We conclude that the bulk of ionized baryonic
matter in the IGM has not been greatly diminished from the epoch
of reionization until the present epoch.

We have argued that VMSs are responsible for the $P$ inventory and
reionization of the IGM. An important issue is the timescale over
which this is achieved. From the range in $z$ where the
model appears to describe the abundances observed in Ly$\alpha$
systems, we infer that the IGM had already acquired the $P$
inventory before $z\sim 4.6$. This is consistent with the following
considerations of the cosmological epoch during which VMSs could
disperse processed baryonic matter to the general IGM.
The explosion energies of VMSs are $\sim
10^{52}$--$10^{53}$ erg as estimated from the available models of
Heger \& Woosley (2002). Based on the cold dark matter model of
structure formation, dark matter halos corresponding to
$1\,\sigma$ fluctuations and collapsing at $z\sim 4$ or those
corresponding to $2\,\sigma$ fluctuations and collapsing at $z\sim
10$ provide binding energies of $\sim 10^{52}$--$10^{53}$ erg for
the baryonic matter in their potential wells (see Figure 9 in
Barkana \& Loeb 2001). Thus, VMS explosions could readily unbind
baryonic matter that was not in stars from most dark matter halos
formed at $z>4$. We conclude that VMSs dominate both the elemental
production and the dispersal of processed baryonic matter at
$z>4$. Compared with the nearly constant values of
$\Omega(\mbox{\ion{C}{4}})$ and $\Omega(\mbox{\ion{Si}{4}})$ at
$z\sim 2$--4.6, the apparent decrease in these two 
quantities from
$z\sim 4.6$ to $\sim 5.3$ (see Table 2 here or Table 1 in 
Songaila 2001) may be a hint for
cessation of VMS activities at $z\sim 4.6$ although this decrease
may also be due to the incompleteness of the sample of Ly$\alpha$
systems in the highest redshift interval.
Songaila and Cowie (2002) have reported a damped Ly$\alpha$
system with [Fe/H]~$\approx -2.7$ and $z=5.3$. This indicates
that VMS activities may have ceased as early
as $z\sim 5.3$. In any case, the bulk
of baryonic matter should have been enriched with the $P$
inventory, ionized, and dispersed in the IGM when the epoch
of VMS activities ended. The subsequent
reaggregation of baryonic matter into galaxies appears to be very
inefficient as demonstrated by the prevalence of the $P$ inventory
in Ly$\alpha$ systems with $z\sim 0.09$--4.6. When galaxies do
form, it appears that the times of formation vary greatly, as
argued by Wasserburg \& Qian (2000b) based on the large scatter in
[Fe/H] for damped Ly$\alpha$ systems at a given redshift in the
range of $z\approx 1.5$--4.5. The precise epoch of VMS activities
and its influence on structure formation at later times are
important issues that remain to be addressed.

We would like to dedicate this paper to the memory of Fritz Zwicky.
We thank Antoinette Songaila, Todd Tripp, and David Tytler for
generously answering many questions and for guiding us through
their data. This work was supported in part by the Department of Energy
under grants DE-FG02-87ER40328 and DE-FG02-00ER41149, and by NASA
under grant NAG5-10293, Caltech Division Contribution 8780(1091).

\clearpage
\begin{deluxetable}{lrrll}
\tabletypesize{\scriptsize}
\tablecaption{$\Omega_P({\rm E})$ Values as Determined from the $P$
Inventory} 
\tablewidth{0pt}
\tablecolumns{5} 
\tablehead{ \colhead{E}&\colhead{$A_{\rm E}$}&
\colhead{$\log\epsilon_P({\rm E})$}& \colhead{$\Omega_P({\rm
E})$}&\colhead{[E/H]}\\
\colhead{(1)}&\colhead{(2)}&\colhead{(3)}&\colhead{(4)}&
\colhead{(5)} } 
\startdata 
C&12&5.11\tablenotemark{a}&$5.56\times
10^{-8}$&$-3.45$\\ N&14&6.30&$1.00\times 10^{-6}$&$-1.75$\\
O&16&6.60&$2.29\times 10^{-6}$&$-2.33$\\ Na&23&3.34&$1.81\times
10^{-9}$&$-2.99$\\ Mg&24&5.13&$1.16\times 10^{-7}$&$-2.45$\\
Al&27&3.12&$1.28\times 10^{-9}$&$-3.35$\\ Si&28&5.02&$1.05\times
10^{-7}$&$-2.53$\\ Ca&40&3.75&$8.09\times 10^{-9}$&$-2.61$\\
Sc&45&0.28&$3.08\times 10^{-12}$&$-2.82$\\ Ti&48&2.43&$4.65\times
10^{-10}$&$-2.56$\\ V&51&1.15&$2.59\times 10^{-11}$&$-2.85$\\
Cr&52&2.40&$4.70\times 10^{-10}$&$-3.27$\\ Mn&55&1.90&$1.57\times
10^{-10}$&$-3.49$\\ Fe&56&4.51&$6.52\times 10^{-8}$&$-3.00$\\
Co&59&2.24&$3.69\times 10^{-10}$&$-2.68$\\ Ni&58&3.25&$3.71\times
10^{-9}$&$-3.00$\\ Cu&63&0.53&$7.68\times 10^{-12}$&$-3.68$\\
Zn&64&1.86&$1.67\times 10^{-10}$&$-2.74$\\ Sr&88&0.13&$4.27\times
10^{-12}$&$-2.77$\\ Ba&138&$-1.80$&$7.87\times 10^{-14}$&$-3.93$\\
\enddata
\tablecomments{Column (2) gives the approximate atomic
mass number for the element; (3) the $P$ inventory; (4) the
fraction of the critical mass density contributed by the element
at the present epoch; and (5) ${\rm [E/H]}=\log\epsilon({\rm E})-
\log\epsilon_\odot({\rm E})$.}
\tablenotetext{a}{The $\log\epsilon_P$ value for C is obtained from 
data on HD 115444 and HD 122563 (Westin et al. 2000).}
\end{deluxetable}

\clearpage
\begin{deluxetable}{lrrrrr}
\tabletypesize{\scriptsize}
\tablecaption{Ionization Fractions for Ly$\alpha$ Systems}
\tablewidth{0pt} 
\tablecolumns{6} 
\tablehead{
\colhead{E}&\colhead{$\Omega({\rm E})$}& \colhead{E$_{\rm
ion}$}&\colhead{$\langle z\rangle$}& \colhead{$\Omega({\rm E}_{\rm
ion})$}& \colhead{$\Omega({\rm E}_{\rm ion})$/$\Omega_P({\rm E})$}\\ 
\colhead{(1)}&\colhead{(2)}&\colhead{(3)}&\colhead{(4)}&
\colhead{(5)}&\colhead{(6)} } 
\startdata 
C&$5.56\times
10^{-8}$&C IV&1.87&$2.05\times 10^{-8}$&0.37\\
&&&2.27&$3.74\times 10^{-8}$&0.67\\ &&&2.78&$2.62\times
10^{-8}$&0.47\\ &&&3.21&$2.46\times 10^{-8}$&0.44\\
&&&3.75&$3.68\times 10^{-8}$&0.66\\ &&&4.24&$3.93\times
10^{-8}$&0.71\\ &&&4.66&$1.35\times 10^{-8}$&0.24\\
&&&5.29&$4.93\times 10^{-9}$&0.09\\ O&$2.29\times
10^{-6}$&O VI&0.09&$4.6\times 10^{-7}$&0.20\\ &&&$\approx
0.24$&$1\times 10^{-6}$&0.44\\ &&&0.9&$\geq 1.3\times
10^{-7}$&$\geq 0.06$\\ &&&$\approx 2.5$&$2.96\times
10^{-7}$&0.13\\ Si&$1.05\times
10^{-7}$&Si IV&2.24&$9.04\times 10^{-9}$&0.09\\
&&&2.81&$1.59\times 10^{-8}$&0.15\\ &&&3.19&$6.19\times
10^{-9}$&0.06\\ &&&3.70&$1.59\times 10^{-8}$&0.15\\
&&&4.28&$1.63\times 10^{-8}$&0.16\\ &&&4.64&$4.66\times
10^{-9}$&0.04\\ &&&5.29&$9.31\times 10^{-10}$&0.01\\
\enddata
\tablecomments{Data on C IV and Si IV from Songaila
2001 and data on O VI from Savage et al. 2002 ($\langle
z\rangle=0.09$), Tripp et al. 2000 ($\langle z\rangle\approx
0.24$), Burles \& Tytler 1996 ($\langle z\rangle=0.9$), and
Simcoe, Sargent, \& Rauch 2002 ($\langle z\rangle\approx 2.5$).
The $\Omega({\rm E}_{\rm ion})$ values in column (5) are 
recalculated from the published data so that these values correspond to 
a flat cosmological model with $\Omega_m=0.3$, $\Omega_\Lambda=0.7$,
and $h=0.65$. Column (2) gives the fraction of the critical mass
density contributed by the element; (4) the average redshift of
the Ly$\alpha$ systems from which column (5) is obtained;  (5) the
fraction of the critical mass density contributed by the ions of
the element; and (6) the ionization fraction $f({\rm E}_{\rm
ion})=\Omega({\rm E}_{\rm ion})/\Omega_P({\rm E})$ estimated
from the $P$ inventory.}
\end{deluxetable}

\clearpage
\begin{references}
\reference{}
Bahcall, J. N., \& Peebles, P. J. E. 1969, \apj, 156, L7
\reference{}
Barkana, R., \& Loeb, A. 2001, Phys. Rep., 349, 125
\reference{}
Boesgaard, A. M., King, J. R., Deliyannis, C. P., \& Vogt, S. S. 1999,
\aj, 117, 492
\reference{}
Bromm, V., Ferrara, A., Coppi, P. S., \& Larson, R. B. 2001,
\mnras, 328, 969
\reference{}
Bromm, V., Kudritzki, R. P., \& Loeb, A. 2001, \apj, 552, 464
\reference{}
Burles, S., Nollett, K. M., \& Turner, M. S. 2001, \apj, 552, L1
\reference{}
Burles, S., \& Tytler, D. 1996, \apj, 460, 584
\reference{}
Burris, D. L., Pilachowski, C. A., Armandroff, T. E., Sneden, C. 2000,
\apj, 544, 302
\reference{}
de Bernardis, P., et al. 2002, \apj, 564, 559
\reference{}
Heger, A., \& Woosley, S. E. 2002, \apj, in press
\reference{}
Israelian, G., Garc\'ia L\'opez, R. J., \& Rebolo, R. 1998,
\apj, 507, 805
\reference{}
McWilliam, A., Preston, G. W., Sneden, C., \& Searle, L. 1995, \aj,
109, 2757
\reference{}
Oh, S. P., Nollett, K. M., Madau, P., \& Wasserburg, G. J. 2001,
\apjl, 562, L1
\reference{}
Prochaska, J. X., Gawiser, E., \& Wolfe, A. M. 2001, \apj, 552, 99
\reference{}
Prochaska, J. X., \& Wolfe, A. M. 2000, \apjl, 533, L5
\reference{}
Qian, Y.-Z., \& Wasserburg, G. J. 2001a, \apj, 549, 337
\reference{}
Qian, Y.-Z., \& Wasserburg, G. J. 2001b, \apj, 559, 925
\reference{}
Qian, Y.-Z., \& Wasserburg, G. J. 2002, \apj, in press
\reference{}
Savage, B. D., Sembach, K. R., Tripp, T. M., \& Richter, P. 2002,
\apj, 564, 631
\reference{}
Simcoe, Sargent, W. L. W., \& Rauch, M. 2002, \apj, submitted
\reference{}
Songaila, A. 2001, \apjl, 561, L153
\reference{}
Songaila, A., \& Cowie, L. L. 2002, \aj, in press
\reference{}
Tripp, T. M., Savage, B. D., \& Jenkins, E. B. 2000, \apj, 534, L1
\reference{}
Wasserburg, G. J., \& Qian, Y.-Z. 2000a, \apjl, 529, L21
\reference{}
Wasserburg, G. J., \& Qian, Y.-Z. 2000b, \apjl, 538, L99
\reference{}
Westin, J., Sneden, C., Gustafsson, B., \& Cowan, J. J. 2000, 
\apj, 530, 783
\end{references}
\end{document}